\documentclass[graybox
]{svmult}

\usepackage{type1cm}        
\usepackage{makeidx}         
\usepackage{graphicx}        
\usepackage{multicol}        
\usepackage[bottom]{footmisc}
\usepackage{hyperref}
\usepackage{extarrows}

\hypersetup{colorlinks=true, linkcolor=black, urlcolor=black, citecolor=blue}

\usepackage{newtxtext}       %
\usepackage[varvw]{newtxmath}       

\makeindex             

\usepackage{tikz}
\usetikzlibrary{fit}
\usetikzlibrary{shapes.geometric}
\usetikzlibrary{decorations.pathmorphing}

\tikzset{
point/.style={circle,fill=black,inner sep=1pt},
vertex/.style={circle,fill=black,inner sep=1.5pt},   
bvertex/.style={circle,fill=black,inner sep=2.8pt},
Bvertex/.style={circle,fill=black,inner sep=4pt}, 
specialEP/.style={rectangle,fill=white,draw,inner sep=3pt},  
whitevex/.style={circle,fill=white,draw, inner sep=2pt},
linelabel/.style={sloped,above,very near start, inner sep=1pt,execute at begin node=$\scriptstyle,execute at end node=$},
baseline=(current  bounding  box.center),doubled/.style={double distance= 1pt,line width=1.5pt},
th/.style={line width=0.5 pt, gray},  
med/.style={line width=1 pt}  
}

\xdefinecolor{myred}{rgb}{0.7,0,0.2} 
\xdefinecolor{mypurple}{rgb}{0.6,0.2,0.4} 
\xdefinecolor{myblue}{rgb}{0.259,0.2,0.6}   
\xdefinecolor{myorange}{rgb}{1,0.6,0.2}   
\definecolor{orange}{rgb}{1,0.5,0}
\xdefinecolor{purpleish}{cmyk}{0.75,0.75,0,0}

\def\bR{\mathbb{R}}

\def\bZ{\mathbb{Z}}
\def\cC{\mathcal{C}}

\def\cO{\mathcal{O}}

\def\cL{\mathcal{L}}

\def\cT{\mathcal{T}}
\def\eps{\varepsilon}
\def\ph{\varphi}

\def\indic{\hbox{\raise-2pt \hbox{\indbf 1}}}

\let\dpr=\partial

\let\==\equiv

\let\io=\infty
\let\0=\noindent

\def\*{{\hfill\break\null\hfill\break}}

\def\ie{\hbox{\it i.e.\ }}
\def\eg{\hbox{\it e.g.\ }}

\def\tende#1{\,\vtop{\ialign{##\crcr\rightarrowfill\crcr
             \noalign{\kern-1pt\nointerlineskip}
             \hskip3.pt${\scriptstyle #1}$\hskip3.pt\crcr}}\,}
\def\otto{\,{\kern-1.truept\leftarrow\kern-5.truept\to\kern-1.truept}\,}

\def \aa{{\mathfrak a}}

\def\be{\begin{equation}}
\def\ee{\end{equation}}

\let\a=\alpha \let\b=\beta    \let\g=\gamma     \let\d=\delta     
             \let\l=\lambda
\let\m=\mu                          \let\r=\rho
 \let\t=\tau         \let\ph=\varphi   
   \let\o=\omega     
 \let\D=\Delta       \let\L=\Lambda

\definecolor{lightblue}{rgb}{0, 0.33, 0.71}

\DeclareMathOperator{\supp}{Supp}

\def\aa{\mathfrak{a}}
\def\bb{\mathfrak{b}}

\def \blue#1 {\textcolor{blue}{#1}}
\def \red#1 {\textcolor{red}{#1}}

\def \blue#1{\textcolor{blue}{#1}}
\def \red#1{\textcolor{red}{#1}}

\begin{document}

\titlerunning{Energy expansions for dilute Bose gases from local condensation results}

\title*{
Energy expansions for dilute Bose gases from local condensation results: a review of known results
}
\author{Giulia Basti, Cristina Caraci  and Serena Cenatiempo}
\institute{Giulia Basti \at  Gran Sasso Science Institute, \email{giulia.basti@gssi.it}
\and Cristina Caraci \at  University of Zurich, \email{cristina.caraci@math.uzh.ch}
\and
Serena Cenatiempo \at  Gran Sasso Science Institute, \email{serena.cenatiempo@gssi.it}}

\maketitle

\abstract{Non-relativistic interacting bosons at zero temperature, in two and three dimensions, are expected to exhibit a fascinating critical phase, famously known as condensate phase. Even though a proof of Bose-Einstein condensation in the thermodynamic limit is still beyond reach of the current available methods, in the past decades the mathematical physics community has gained an enhanced comprehension of other aspects of the macroscopic behavior of dilute Bose gases at zero temperature. In these notes we review part of these advances, by focusing on the strict relation among the occurrence of Bose-Einstein condensation 
on large -- but finite -- boxes, and the asymptotic expansion to the ground state energy of dilute Bose gases.}

\section{Introduction} \label{sec:intro}

We consider $N$ bosons in the $d$-dimensional torus $\L_T = [-T/2, T/2]^d$, $d=2,3$,  interacting via a two body non negative, radial and compactly supported potential $V$ with scattering length $\aa$ (see \cite[App.C]{LSSY} for a definition of the scattering length). Note that with a slight abuse of notation we denote by $\aa$ both the three and two dimensional scattering lengths. In the units where the particle mass is set to $m=1/2$ and $\hbar=1$, the Hamilton operator has the form
\be \label{eq:HT}
H_T = -\sum_{i=1}^N \D_{x_i} + \sum_{1\leq i<j \leq N} V(x_i-x_j)
\ee
and acts on the Hilbert space $L^2_s(\L_T^N)$, the subspace of $L^2(\L_T^N)$ consisting of functions which are symmetric with respect to permutations of the $N$ particles. Here the letter $T$ for the side length of $\L$ stands for `thermodynamic box', since we are going soon to consider the limit $N,T\to \io$ with $\r=N/|\L_T|$ fixed, known as {\it thermodynamic limit}.

In absence of interaction, and in the thermodynamic limit, the system described by \eqref{eq:HT} exhibit a striking feature: at sufficiently low temperature in three dimensions and at zero temperature in two dimensions  a macroscopic fraction of the particles (depending on the temperature) occupies the ground state of the kinetic energy operator, namely the zero momentum mode (see \eg \cite[Sec.1]{LSSY}).  
This macroscopic behaviour, also known as {\it Bose-Einstein condensation} (BEC),  is expected to survive for interacting systems (at least for $\r \aa^d$ small enough),
at low temperature in three dimensions and at zero temperature in two dimensions (condensation at positive temperature for interacting bosons in two dimensions is ruled out by the Hohenber-Mermin-Wagner theorem \cite{H,MW}).
Still a proof of the occurrence of BEC for generic bosonic systems is out of reach of rigorous analysis (see \cite{DLS, KLS} for the only available result to date). This fact is not surprising, since  the emergence of BEC is an example of occurrence of long-range order in systems with continuous symmetries, see e.g. \cite[Sect. 2]{SP-book}, \cite{LSY-BEC-SSB} and \cite[Ch. 5]{Tasaki}.

In these notes we are interested in a different - though related - aspect: providing asymptotic expressions for the thermodynamic functions of dilute Bose gases in the thermodynamic limit. As an instance of this goal, we will consider the ground state energy per unit volume in the thermodynamic limit. 
Let us denote by $E_d(N,T)$ the ground state energy of the system described by \eqref{eq:HT} in $d=2,3$ dimensions. Then the specific ground state energy, defined by
\be\label{eq:e_d}
e_d(\r) = \lim_{\substack{N, T \to \io\\ \r=N/T^d}} \frac{E_d(N,T)}{T^d}
\ee
admits the following expansions in the dilute limit $\r \aa^d \ll 1$: 
\begin{align} 
e_{3}(\rho) &\;= 4\pi \rho^2 \aa  \Big[1+\frac{128}{15\sqrt{\pi}}\sqrt{\rho \aa^3}+o\big(\sqrt{\rho \aa^3}\big)\Big]\,, \label{eq:LHY} 
 \\
e_2(\r) &\;= 4 \pi \r^2 \, \bb \Big[ 1 - \bb  | \log \bb | +  \Big( \frac 12 + 2\g + \log \pi \Big)  \bb + o(\bb)\Big] \label{eq:LHY2d}\,,
\end{align}
with $\bb= |\log (\r \aa^2)|^{-1}$. 
The expansions \eqref{eq:LHY} and \eqref{eq:LHY2d}, first predicted in \cite{B,LHY} for the $3d$ case, and in \cite{Bog2D-5,A2d,MC03} for the $2d$ case, have been recently proven to be correct for non negative interactions with finite scattering length, see \cite{FS, FS2, YY, BCS} and  \cite{FGJMO} respectively. 
More precisely, lower and upper bounds compatible with \eqref{eq:LHY2d} have been proved in \cite{FGJMO} for generic non negative potentials, including the hard core case. Differently, the conditions under which the correct three dimensional upper bound can be proven are much more restrictive ($V \in L^3$) than those necessary for the lower bound, where the hard core case is included. We will come back to the difference among the $2d$ and $3d$ upper bounds in Section \ref{sec:UB}.

A striking feature of \eqref{eq:LHY} is its universal behaviour: in the dilute regime the ground state energy up to second order does not depend on the detail of the system in consideration, but only on the particle density and the scattering length of the interaction. 
These physical quantities also define the relevant length scales for the problem, which are:  the scattering length $\aa$ of the interaction; the mean inter-particle distance $\r^{-1/d}$; the inverse square root of the ground state energy per particle, given by $(\r \aa)^{-1/2}$ in three dimensions and $(\r \bb)^{-1/2}$ in two dimensions. The last length scale, usually referred to as {\it de Broglie wavelength}, {\it uncertainty principle length}, or {\it healing length},
defines the length below which a particle cannot be localized without seriously altering its energy (see the remark after Eq.\eqref{eq:localiz-UB} in Sec. \ref{sec:UB}).
In the dilute regime, we have the relations: 
\[  \begin{split}
\aa & \ll \r^{-1/3} \ll (\r \aa)^{-1/2} \qquad d=3 \\
\aa & \ll \r^{-1/2} \ll (\r \bb)^{-1/2} \qquad d=2 \,.
\end{split}\]
Finally, let us mention a last important length scale, which is the one identified by the range of the interaction $R \geq \aa$. In particular it enters in the definitions of  {\it high density} and {\it low density} regimes, corresponding to situations where $\r R^d \gg 1$ and $\r R^d \ll 1$ respectively. 
Note that a system can be both dilute $\r \aa^d \ll 1$ and high density, provided that $\aa/R \ll 1$. 

We will see in the next sections that the length scales introduced above play a key role in the methods allowing for a rigorous derivation of the expansions \eqref{eq:LHY} and \eqref{eq:LHY2d}. Note that in these notes we will not discuss the dilute limit for bosonic systems in one dimension, which is very different from the two and three dimensional cases considered here.  We refer the reader to \cite{ARS}, where the latest result in this setting has been recently obtained. Moreover we will only discuss the zero temperature phase; recent results on the macroscopic properties of dilute bosons at positive temperature and in the thermodynamic limit have been obtained in \cite{Sei-T, Yin, NRS2}.

\medskip

 {\bf Bogoliubov theory.} The first  understanding of the expansions in \eqref{eq:LHY}
 and \eqref{eq:LHY2d}  
 is based on an approximate exactly solvable model due to Bogoliubov \cite{B} (see also \cite[Appendix A]{LSSY} for a review).  Bogoliubov started from the assumption (so far not yet justified) that low-energy states of \eqref{eq:HT} exhibit Bose-Einstein condensation for sufficiently weak interaction. 
Guided by this idea, he rewrote the Hamilton operator \eqref{eq:HT} in momentum space, using the formalism of second quantization, and replaced all creation and annihilation operators associated with the zero-momentum mode  by factors  of $N^{1/2}$. The resulting Hamiltonian contains a constant term (describing the interaction among particles in the condensate), plus terms containing creation and annihilation operators associated with modes with momentum $p \not = 0$, usually referred to as {\it excitations}. In particular we have:
 i) terms that are quadratic in creation and annihilation operators associated with $p \not = 0$ modes, describing the kinetic energy of the excitations as well as the interaction among pairs of excitations and the condensate; ii) terms that are cubic and quartic, describing interactions involving more than two excitations. Neglecting all cubic and quartic contributions, Bogoliubov obtained a quadratic Hamiltonian that he could diagonalize explicitly,  obtaining the following expression for the ground state energy of the system of \eqref{eq:HT}:
 \be  \label{eq:BogEnergy-anyd}
 \begin{split}
E_d(N,T) =\; & \frac{N }{2} \rho \widehat V(0) - \frac 1 2 \sum_{\substack{p \in \frac{2\pi}{T} \bZ^d,\\ p \neq 0}} \Big[\; p^2 + \rho \widehat V(p) - \sqrt{p^4  \hskip -0.05cm+  \hskip -0.05cm2 \rho \widehat V(p) p^2 }\;  \Big] \,, 
\end{split}\ee
where $\widehat V(p)$ denotes the Fourier transform of $V$, defined as
\be \label{eq:Fourier}
\widehat V(p)= \int_{\L_T} e^{ip\cdot x} V(x) d^d x\,.
\ee
Let us now focus on the three dimensional case, which is the one directly analysed in \cite{B}. We rewrite  \eqref{eq:BogEnergy-anyd} as follows 
\be  \label{eq:BogEnergy} \begin{split}
E_3(N,T) =\; & \frac{N }{2} \rho \widehat V(0) - \frac 1 4 \hskip -0.1cm\sum_{\substack{p \in \frac{2\pi}{T} \bZ^3 \\ p \neq 0}} \frac{ (\rho \widehat V(p) )^2 }{p^2} \\
&- \frac 1 2 \sum_{\substack{p \in \frac{2\pi}{T} \bZ^3 \\ p \neq 0}} \Big[ p^2 + \rho \widehat V(p) - \sqrt{p^4  \hskip -0.05cm+  \hskip -0.05cm2 \rho \widehat V(p) p^2 }  -  \frac 1 2  \frac{ (\rho \widehat V(p) )^2 }{p^2} \Big] \,.
\end{split}\ee
and then we take the thermodynamic limit $N,T \to \io$, $\r=N/T^3$. 
One easily recognizes, in the expressions appearing on the r.h.s. of  \eqref{eq:BogEnergy}, the first and second term of the Born approximations of the scattering length $\aa$ of $V$,  which reads:
\[
\aa = a_0 + a_1 + \cO\big((a_0/R)^2\big) \,,
\]
with
\be \label{eq:Born}
8 \pi a_0=  \widehat V(0)\,, \qquad   8 \pi a_1= -  \int \frac{d^3 p}{(2\pi)^3} \frac{\widehat V(p)^2}{2p^2} \,. \ee
Let us recall that $\aa$ is defined through the solution $f$ of the zero energy scattering equation, which in three dimension solves
\be\label{eq:scatt_3d}
    \Big(-\D + \frac12 V\Big)f =0\,, \qquad f(x) \xlongrightarrow[x \to \io]{} 1\,.
\ee
With this definition we have $\int V(x)f(x)\,d^3x=8\pi\aa$.

Eq.\,\eqref{eq:LHY} can be then obtained by replacing the sum $a_0 + a_1$  by  $\aa $ in  first line of Eq.~\eqref{eq:BogEnergy}, and $\widehat V(p)$ by $\widehat V(0) \simeq 8 \pi \aa $ in the integral obtained from the sum on the second line of the same equation\footnote{In fact, in his paper Bogoliubov first derived \eqref{eq:BogEnergy-anyd} (see \cite[Eq.\,(16)]{B} and then suggested that the general expression for the ground state energy of dilute bosons could be obtained by replacing in~\eqref{eq:BogEnergy-anyd} the Fourier coefficients $\widehat V(p)$ by $\widehat{(V \ast f)}(p)$, see the comment around \cite[Eqs.\,(30)-(31)]{B}. }.
Note that the replacement of $\widehat V(p)$ with $\widehat V(0)$ in the integral is possible thanks to the fact that we have added to the sum on \eqref{eq:BogEnergy-anyd} the term $ \sum_{ p \neq 0} (\r\widehat V(p))^2/ (4 p^{2})$, which in the thermodynamic limits reproduces $- 4 \pi a_1 N \r $.  Let us now denote by $\r_+$ the density of particles outside the condensate in the ground state (the so called {\it condensate depletion}). Then Bogoliubov theory predicts (see \eg \cite[Eq.(4.48)]{SP-book}):
\be \label{eq:depletion} 
\frac{\rho_+}{\r}=  \frac{8}{3 \sqrt \pi}  \sqrt{\r a_0^3} \simeq \frac{8}{3 \sqrt \pi}  \sqrt{\r \aa^3}\,,
\ee
with the usual replacement $a_0 \to \aa$.

\medskip

Let us now analyse Eq.\eqref{eq:BogEnergy-anyd} in the two dimensional case. We recall that for a fixed $\ell>R$, the zero energy scattering function in two dimensions solves
\be \label{eq:scatt_2d}
\Big( -\D + \frac 12 V \Big) f_\ell =0\,,\qquad f_\ell(x)\big|_{|x|=\ell} =1 \,.
\ee
In particular 
\[
f_\ell (x) = \frac{\log(|x|/\aa)}{\log(\ell/\aa)}\,, \qquad \forall x : R <|x| \leq \ell
\]
with $\aa$ denoting here the two dimensional  scattering length. 
For small interactions $V(x)=\l \widetilde V(x)$, with $0<\l \ll 1$, the Born series for the two dimensional scattering length reads:
\[ \label{eq:bell}
|\log (\ell/\aa)|^{-1}  = (4 \pi)^{-1} \widehat V(0) + \sum_{k=1}^{\io} (4\pi)^{-(k+1)} \int_{\bR^2} \big(\cL_V\big)^{k}(V)(x) dx = \sum_{k=0}^\io  2 b_k^{(\ell)}\,,
\]
where $\cL_V$ is the operator given by $\cL_V(g)(x)= V(x)\int_{\bR^2} \log(|x-y|/\ell) g(y)dy$. Clearly  $b_k^{(\ell)}= \cO(\l^{k+1})$. Hence, setting $\ell= \r^{-1/2}$ we obtain the following expansion for the effective parameter $\bb=|\log(\r \aa^2)|^{-1}$ appearing in \eqref{eq:LHY2d}: 
\be \label{eq:bb-Born}
\bb = b_0 + b_1 + b_2 + \cO(\l^4)\,, \qquad \text{where } b_k :=  b_k^{(\r^{-1/2})}\,.
\ee
In particular $ 8 \pi b_0 = \widehat V(0)$ and 
\[
8 \pi b_1 =  \frac{1}{4\pi} \int V(x) V(y) \log(|x-y|/\ell) \, dx dy \,.
\]
Note that with the choice of $\ell$ above the following is satisfied: $R \ll \ell \ll (\r \bb)^{-1/2}= \cO\big((\l \r )^{-1/2}\big)$, \ie $\ell$ is smaller than the $2d$ healing length.
 Similarly to what is done in the three dimensional case we can add and subtract on the r.h.s. of \eqref{eq:BogEnergy-anyd} a term of the form $ \frac {\,\r^2} 2 \sum_{p \neq 0} g_p$, with $g_p$ chosen so that $\frac {\,\r^2} 2 \sum_{p \neq 0}  g_p$ converges to $8 \pi b_1 N \rho$ in the limit $T\to \io$ with $\r=N/T^2$ fixed\footnote{Note that due to the logarithmic divergence of the two dimensional Green function, the expression for $g_p$ is slightly more complicated than in the three dimensional case (see for example \cite[App. A]{FNRS-2d} or \cite[Eq. after (4.17)]{CCS2}), and in particular $g_p \to 0$ in the limit $p \to 0$.}.
Eq. \eqref{eq:LHY2d} is then obtained by taking the thermodynamic limit, and by substituting in the final result:  $ 4 \pi N \r (b_0+b_1)$ with  $ 4 \pi N \r \bb$; $b_0$ with $\bb$ in the terms of the order $b_0^2 \log b_0$ and $b_0^2$ coming from the limit of the sum of the second term on the r.h.s. of  \eqref{eq:BogEnergy-anyd} and  $- \frac {\, \r^2} 2 \sum_{p \neq 0} g_p$. One can also evaluate the condensate depletion, which Bogoliubov's approach predicts to be proportional to $\r \bb$.
We refer to \cite{MC09} and references therein for the use of Bogoliubov's method in two dimensions in the physics literature. In the next paragraphs we will rather discuss the validity of Bogoliubov's approach, both in three and two dimensions, from a mathematical point of view.

\bigskip

{\bf Validity of Bogoliubov's approach in $3d$.} The discussion around \eqref{eq:Born} makes apparent that there exists a regime of parameters $\{R,\aa,\r \}$ where the approximations made by Bogoliubov can be proven to be correct. 
This is in fact the case when $a_0 = (8\pi)^{-1}\widehat V(0)\ll R$ and moreover
\be \label{eq:Bog-exact}
\frac{a_0}{R}  \gg \sqrt{\rho a_0^3} \gg \frac{a_0^2}{R^2}\,.
\ee
The above conditions guarantee that the final replacements made below \eqref{eq:scatt_3d} 
 to obtain \eqref{eq:LHY} only produce subleading contributions (see also the remarks at the end of \cite[Section 1]{BriS}). In other words, whenever \eqref{eq:Bog-exact} holds,  the contribution to the ground state energy coming from the cubic and quartic terms neglected in Bogoliubov theory is negligible at the order of precision we are looking at. 

\begin{figure}[t]
\centering
\begin{tikzpicture}[scale =0.7,
    every path/.style = {},
 ]
  \begin{scope}
  \draw[|-, thick] (0,0)--(3,0);
  \draw[|-, thick] (3,0)--(4,0);
    \draw[|-|, color=black, thick] (4,0)--(6,0);
 \shade[top color=lightgray, bottom color=white, opacity=0.8]    (3.01,-0.25) rectangle (5.99,0.25);
  \draw[<->, thick, black] (4,0.7)--(6,0.7);
  \node[below, text width=8.5em, align=center] at (5,1.5) {\textcolor{black}{\footnotesize high density regime}};
\node[below] at (0, -0.2) {$0$};
\node[below] at (3, -0.2) {$\frac 1 4$};
\node[below] at (4, -0.2) {\textcolor{black}{$\frac 1 3$}};
\node[below] at (6, -0.2) {$\frac 1 2$};
\node[above] at (0.5, 0.2) {$\g$};
\draw[->] (18/5,-1) --(18/5,-0.4);
\node[below] at (18/5, -1){\cite{BriS}};
\draw[->] (134/23,-1) --(134/23,-0.4);
\node[below] at (134/23, -1){\cite{GiuS}};
\end{scope}
\end{tikzpicture}
\caption{For $\r \aa^3 \ll 1$ (dilute regime), the figure above depicts the range of validity of Bogoliubov's approximation in $3d$ (shadowed area) with respect to the value of the parameter $\g>0$ defining the ratio $a_0/R=(\r \aa^3)^\g$. Note that for $\g>1/3$ the systems is in a high density regime, namely $\r R^3 \gg 1$. The case $\g=0$ corresponds to the case where no Born series for the scattering length is available. The threshold reached by \cite{GiuS} ($\g>1/2 - 1/69$) and \cite{BriS} ($\g>3/10$) are also indicated.} 
\label{Fig:Bog}
\end{figure}
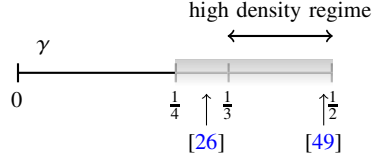

To make the conditions \eqref{eq:Bog-exact} more explicit, 
let us introduce the parameter $\g>0$ defined by $a_0/R = (\r \aa^3)^\g$ (recall that $\r \aa^3 \ll 1$). Then the conditions in \eqref{eq:Bog-exact} are satisfied for $\g \in (1/4, 1/2)$, see the shadowed area in Fig.\ref{Fig:Bog}. On the other hand for $\g \in (1/3,1/2]$ we have
\[ \label{eq:highdensity}
\r R^3  \sim (\r \aa^3)^{1-3\g} \gg 1 \,,
\]
namely the system is in a high density regime.
Indeed, mathematically, the validity of Bogoliubov's approach for three-dimensional Bose gases has been  at  first established in weak coupling and high density regimes:  in \cite{LS-jellium} where the ground state energy of bosonic jellium was obtained, in \cite{LS-charged,S-charged} where a similar result was achieved in the context of the two-component charged Bose gas, and later by Giuliani and Seiringer for the computation of \eqref{eq:LHY}, as discussed in the Sect. \ref{sec:LB} below.

\smallskip

\medskip

The first paper where the validity of Bogoliubov's predictions were proven in a region of parameters where \eqref{eq:Bog-exact} fails is due to Erd\H os, Schlein and Yau in \cite{ESY}.
There, an upper bound for the specific ground state energy was proven considering an interaction of the form $\lambda V$ for $\lambda>0$ small. The result coincides with the asymptotic in Eq.\eqref{eq:LHY}, except for the fact that the second order correction is multiplied by a constant $S_\lambda$ satisfying $1\leq S_\lambda\leq 1+C\lambda$. Note that a trial state as in \cite {ESY} provides  an upper bound in agreement with \eqref{eq:LHY} under the assumption  $a_0/R=(\r \aa^3)^\gamma$ for $\gamma>0$ even though \eqref{eq:Bog-exact} fails as soon as $\gamma\leq1/4$.
The key idea in \cite{ESY} is that a good trial state capturing \eqref{eq:LHY} for smooth potentials is given by a quasi-free state on the top of the condensate, with the quasi-free state describing the creation of couples of excited particles with momenta $(p,-p)$. In \cite{ESY} the creation of couples of excitations is mediated by a kernel $c_p$ which has quite a different behaviour for momenta smaller or higher than the inverse healing length $(\r \aa)^{1/2}$. Let $f$ be the solution of the zero energy scattering equation defined in \eqref{eq:scatt_3d}. Then for high momenta $c_p$ is the Fourier transform of $\r (1-f)$ and for low momenta is a quantity related to the diagonalization of a quadratic Hamiltonian, which can be thought as obtained by conjugating the Bogoliubov's pair Hamiltonian (as in \cite[A.17]{LSSY}) through a Bogoliubov's transformation with kernel $\widehat{\rho(1-f)}_p$. 

An important remark is that the result from \cite{ESY} do not improve with different choices of the coefficients $c_p$. Indeed, as showed in \cite{NRS1,NRS2} quasi-free states can only approximate the specific ground
state energy of a $3d$ dilute Bose gas up to errors which are of order $\r^2\aa \sqrt{\r \aa^3}$. To show \eqref{eq:LHY} for general dilute systems, it is needed to go beyond quasi-free states, in order to be able to extract some missing contribution hidden in the cubic and quartic terms \cite{YY, BCS, FS,FS2}. 

Finally, let us mention that a  systematic study of the perturbation theory around Bogoliubov's model in three dimension, and the proof of its order by order convergence after proper resummations, was obtained by Benfatto back in 1997 \cite{Benfatto} (see \cite[Ch. 1]{PhD-Cenatiempo} for an outline of this approach). However the resulting bounds are not enough for constructing the theory in a mathematically complete
form: they allow to derive finite bounds at all orders in renormalized perturbation
theory, growing like $n!$ at the $n$-th order, but the possible Borel summability of the series remains an outstanding open problem. A program addressing this issue has been
started by Balaban, Feldman, 
Kn\"orrer and Trubowitz, see \cite{Balaban} for an overview.

\bigskip

{\bf Validity of Bogoliubov's approach in $2d$.} As commented around \eqref{eq:bb-Born}, in two dimensions Bogoliubov's approach is expected to give the correct asymptotics for sufficiently weak interactions, so that  
\be\label{eq:Bog-exact2d}
 b_2 \ll \max\{ b_0, b_1, b_0^2, b_0^2 \log  b_0\}\,.
\ee
Note that, differently from the three dimensional case, the range of the interaction does not enter in the identification of the regimes in \eqref{eq:Bog-exact2d}. Still we can expect Bogoliubov theory to hold both for low density and high density regimes since for interactions of intensity $\l \ll 1$ we have $R \ll ( \l \r)^{-1/2}$, that is  $\rho R^2 \ll 1/\l$.

The validity of  Bogoliubov's two term asymptotics for the ground state energy  for sufficiently weak interactions has been indeed shown in \cite{FNRS-2d}, by considering quasi-free states. 
Very recently, a proof establishing \eqref{eq:LHY2d} for generic non negative potentials and dilute $2d$ systems has been obtained in \cite{FGJMO} (see Sec. \ref{sec:LB} and Sec. \ref{sec:UB} below for more details).  
Finally, in the case of weak interactions, a systematic study of the corrections to Bogoliubov theory
via renormalization group methods (allowing in principle also to obtain information on correlation functions) can be obtained up to length scales of the order $\l^{-1} (\l \r)^{-1/2} $, which is the largest scale at which the problem has a perturbative nature \cite{CG}.

\medskip

\medskip

{\bf Summarizing.} The fact that the expansion \eqref{eq:LHY} (resp.\eqref{eq:LHY2d}) has been shown to be correct in more general regimes than those identified by \eqref{eq:Bog-exact} (resp. \eqref{eq:Bog-exact2d}) shows that Bogoliubov's predictions remain corrects beyond the regimes of validity of Bogoliubov's approximations. We will see in the next sections, that this is linked to two main facts: 
\begin{enumerate}
\item[i)] to make Bogoliubov theory rigorous it is sufficient to show the occurrence of condensation on a sufficiently large, but finite  box;  \smallskip
\item[ii)] in situations where \eqref{eq:Bog-exact} (resp. \eqref{eq:Bog-exact2d}) are not satisfied, it is possible to extract from the cubic and quartic terms neglected in Bogoliubov theory, some contributions to the ground state energy which renormalize the interaction and make the scattering length to appear. 
\end{enumerate}
Finally, the problem shows  a clear separation of energies: momenta of the order $\aa^{-1}$ are responsible for the appearance of the full scattering length, while momenta of the order of  $(\rho\aa)^{1/2}$ in $3d$ and $(\rho\bb)^{1/2}$ in $2d$ are responsible for the second term in \eqref{eq:LHY} and \eqref{eq:LHY2d}.

\section{Occurrence of local condensation: scaling regimes } \label{sec:scalings}

A natural question that arises from the overview given in the previous section is what is the largest length scale at which Bose-Einstein condensation can be shown to occur, and whether the results predicted by Bogoliubov, which are in fact based on the occurrence of BEC, can be validated up to the same scale.
The first result in this direction is due to  Lieb and Seiringer who, in the celebrated paper \cite{LS}, exhibited a proof of Bose-Einstein condensation for $3d$ interacting bosons in the so-called Gross-Pitaevskii regime, famously introduced in \cite{LS1,LSY}.

Let us consider $N$ bosons trapped in a region of order $L$, and interacting through a potential whose scattering length $\aa$ is of the same order of the range as the interaction $R$. Then the $3d$ Gross-Pitaevskii regime is defined as the limit where we send $L, N$ to infinity by keeping the ratio $\aa N /L$ fixed.
Note that, if we consider $\aa$ fixed as in the previous section, the Gross-Pitaevskii scaling corresponds to an ultradilute regime where the particle density goes to zero as $N^{-2}$. On the other hand, if we rescale lengths and rather consider $L$ of order one, we have that the Gross-Pitaevskii regime corresponds to consider potentials with a scattering length of order $N^{-1}$ (which are nicely describing the intense and short range interactions typical of BEC in cold atomic gases). In the simplest case where the bosons are trapped on a three dimensional torus the Gross-Pitaevskii Hamiltonian is an operator on $L^2_s([0,1]^3)$ of the form:
\be \label{eq:HN-GP}
H_N^{(GP)}= \sum_{i=1}^N -\D_{x_i} + \sum_{1\leq i < j \leq N} N^{2} V(N (x_i- x_j))\,.
\ee
In \cite{LSY} Lieb, Yngvason and Seiringer were  able to provide an expression for the ground state energy of  $H_N^{(GP)}$ compatible with \eqref{eq:LHY} at leading order, but also valid in the more general case where the bosons are confined in a region of order $L$ by a generic trapping potential (in the latter case the ground state energy is obtained through the minimization of a non linear energy functional on $L^2(\bR^3)$, see \cite[Theorem 6.1]{LSSY}).
In the same years the same results were extended to two dimensions \cite{LS1,LSY2}. There the Gross-Pitaevskii Hamiltonian on $L^2_s([0,1]^2)$ is defined to be:
\be \label{eq:HN-GP2d}
H_{N,2d}^{(GP)}= \sum_{i=1}^N -\D_{x_i} + \sum_{1\leq i < j \leq N} e^{2N} V( e^N (x_i- x_j))\,.
\ee
The exponential scaling of the interaction potential in \eqref{eq:HN-GP2d} is due to the fact that in $2d$ the Gross-Pitaevskii regime is defined as the limit where $N \to \io$ with $\bb N$ fixed, and $\bb$ depends logarithmically on the scattering length.
Note that, rescaling lengths, the two dimensional Gross-Pitaevskii regime can be interpreted as describing an extended Bose gas (of particles interacting through the unscaled
potential $V$) at a density that is exponentially small in $N$. While the exponential smallness of the density (or, equivalently, of the scattering length) makes it difficult to
directly apply the results obtained in this regime to physically relevant situations, the $2d$ Gross-Pitaevskii provides the simplest scaling limit where the strong correlations typical of two dimensional systems drastically reduce the effective coupling $\bb$.

\medskip

The key fact making possible to show the occurrence of condensation in the Gross-Pitaevskii regime is that in this regime the energy gap between the ground state of the kinetic energy and the first excited state is of the same order of magnitude of the energy per particle (hence one can for example apply the Temple inequality, as discussed in Sec. \ref{sec:LB}). Note also that proving BEC in the Gross-Pitaevskii regimes corresponds to showing condensation on a length scale of order $L \sim (\r \aa)^{-1/2}$ in three dimensions and $L\sim (\r \bb)^{-1/2}$ in two dimensions.

However, the Born series expansion for the scattering length fails to be true in the three dimensional Gross-Pitaevskii regime, unless the potential $V$ is of the form $\l V$ with $0<\l \ll 1$ (since otherwise the ratio among the scattering length of the interaction and its radius is of order one). Moreover, it is never valid in the two dimensional Gross-Pitaevskii regime, since in this regime the integral of the potential is independent of $N$, while $\bb \sim N^{-1}$. 
This is the reason why showing the validity of Bogoliubov theory in the Gross-Pitaevskii regimes has required twenty years more \cite{BBCS1,BBCS3,BBCS4, CCS1, CCS2}, with respect to the pioneering works by Lieb, Seiringer and Yngvason (later revised in \cite{NRS}). 
Note that thanks to the presence of a spectral gap, in the Gross-Pitaevskii regime it also possible to analyse the low energy excitation spectrum \cite{BBCS4, CCS2}, which is particularly important for understanding superfluidity.  

We refer to \cite{Schl22} and \cite{Caraci} for a discussion of the methods allowing for a rigorous implementation of Bogoliubov theory in the $3d$ and $2d$ Gross-Pitaevskii regimes respectively. We remark that the validity of Bogoliubov theory can also be established in presence of a generic trapping potentials \cite{BSS,BSS1,NNRT,NT},  and that more recently a simplified approach to verify Bogoliubov theory in the Gross-Pitaevskii regime has been developed in \cite{Hainzl,HST}. Note that the current method used to recover Bogoliubov theory in the Gross-Pitaevskii regime requires the potential to be at least $L^1$;  a second order upper bound valid for hard core potentials in the Gross-Pitaevskii regime has been recently showed in \cite{BCOPS}. Finally, positive temperature properties of interacting bosons in the Gross-Pitaevskii regime have been analyzed in \cite{DSJ,DS}.

\medskip

In the next two paragraphs we see how one can consider less (or more) singular regimes with respect to the Gross-Pitaevskii one. We will analyse the $3d$ and $2d$ cases separately.

\bigskip \medskip

{\bf Three dimensions.} As before let us consider $N$ bosons on the unitary torus. Let $R$ be the range of the interaction. Scaling regimes less singular than the Gross-Pitaevskii regime can be obtained by choosing $\aa \sim N^{-1}$ and $R \sim N^{-\b}$ with $\b \in [0,1)$ so that  $\aa \ll R$, namely we consider
\be \label{eq:HN-beta}
H_{N,3d}^{(\b)}= \sum_{i=1}^N -\D_{x_i} + \sum_{1\leq i < j \leq N} N^{3\b-1} V(N^\b (x_i- x_j)) \qquad \text{on} \qquad L^2_s([0,1]^3)\,.
\ee
Note that the regime corresponding to $\b=0$ is also known as {\it mean-field} scaling, and for  $\b=1$ we recover the Gross-Pitaevskii regime. We remark that: i) for $\b\in[0,1/3)$ we have $\r R^3 \gg 1$ hence the system is in a high density regime; ii) for $\b\in [0,1/2)$ the conditions \eqref{eq:Bog-exact}  holds, and one can show that Bogoliubov's approximation is correct.
For $\b \in [1/2,1)$, despite the fact that the approximations made by Bogoliubov  fail, it is possible to recover Bogoliubov's predictions \cite{BBCS2} considering a quasi-free state which has a different behaviour on low and high energy scales \cite[Sec. 2]{Olgiati}, in the same spirit of what is done in \cite{ESY} for weakly interacting bosons in the thermodynamic limit. For a comparison among the regimes described by \eqref{eq:HN-beta}  and those discussed around \eqref{eq:Bog-exact}, we refer the reader to \cite[Eqs. (6)-(7)]{BriS}.

\medskip

{\it Remark.} The first rigorous proof of the validity of Bogoliubov's predictions for the ground state energy and low energy spectrum of Hamiltonians as in \eqref{eq:HN-beta}, was given by Seiringer for $\b=0$ \cite{Sei}. This result was later extended in various directions in \cite{GS,LNSS,DN,PizzoI, PizzoII, PizzoIII,BPS,Mitrouskas,LNR2, BroS}. Note that the mean field regime can also be regarded as a semiclassical limit, see \cite{AN1,AN2,AN3,AN4,AFP} and references therein. As for the regimes corresponding to $\b \in (0,1)$ the leading order results for the ground state energy and a proof of condensation without optimal rate were first achieved in \cite{LNR}, which also covers the case of attractive interactions in the
regime of stability. Differently, the validity of Bogoliubov predictions (for all $\b \in (0,1)$) have been established so far only for non negative interaction \cite{BBCS2}.

\medskip

Let us now consider regimes more singular than the $3d$ Gross-Pitaevskii, by taking $\aa \sim R$ and letting the product $\aa N$ grow with $N$.  
To describe these regimes we introduce a parameter $\kappa \in [0,2/3)$ and consider the following Hamilton operator on $L^2([0,1]^3)$:
\be \label{eq:HN-kappa}
H_{N,3d}^{(\kappa)}= \sum_{i=1}^N -\D_{x_i} + \sum_{1\leq i < j \leq N} N^{2(1-\kappa)} V(N^{1-\kappa} (x_i- x_j))\,,
\ee
so that $\aa N \sim N^\kappa$. We note that the choice $\kappa=0$ corresponds to the Gross-Pitaevskii regime. On the other hand, by rescaling lengths so that the scattering length is of order one, we see that \eqref{eq:HN-kappa} is equivalent to studying a Hamiltonian as in \eqref{eq:HT} on the torus $[0,L_\kappa]^3$ with $L_\kappa = N^{\kappa-1}$. This corresponds to systems with density $\r_\kappa = N^{3\kappa-2}$, hence less and less dilute as $\kappa \to 2/3$. In particular the choice $\kappa=2/3$ would correspond to the thermodynamic limit. From the point of view of length scales, proving BEC for Hamiltonians described by \eqref{eq:HN-kappa} is equivalent to showing condensation up to length scales of the order 
\be \label{eq:L3d}
L \sim (\r \aa)^{-1/2} (\r \aa^3)^{-\a}\,, \qquad \a=\kappa /(4-6\kappa)\,. 
\ee
Indeed, the analysis carried out in \cite{LS,LS1} allows to show condensation also beyond the Gross-Pitaevskii regime, up to $\kappa<1/10$.  On the other hand, in the recent paper \cite{F}, this result was extended to all $\kappa<2/5$.
As for the validity of Bogoliubov theory, this was established in \cite{BCaS} for sufficiently small $\kappa>0$, 
by exploiting key estimates on the energy and (powers of) the number of excitations obtained in \cite{ABS}. Finally, we mention \cite{BCS} and \cite{Ba} where a second order upper bound on the ground state energy compatible with \eqref{eq:LHY} is proved for all $\kappa<7/12$. 

We will discuss in the next two sections how the analysis of Hamiltonians of the form \eqref{eq:HN-kappa} is closely related to proving lower and upper bounds compatible with \eqref{eq:LHY}. In particular, current methods (which will be discussed below) allow to obtain lower and upper bounds compatible with \eqref{eq:LHY} starting from systems defined on  a torus of side length $L\sim N^{\kappa-1}$, at the expenses of errors which are smaller than $\r^2 \aa \sqrt{\r \aa^3}$ if $\kappa>0$ in the case of lower bounds, and $\kappa>1/2$ for upper bounds.

\bigskip

{\it Thomas-Fermi regime.} A scaling regime which is closely related to the Gross-Pitaevskii one, but will not discussed in these notes is the Thomas-Fermi regime \cite[Ch. 6]{LSSY}. In this regime
one considers a system of interacting bosons in a dilute regime with scattering length $\aa \gg R$ and so that the product $N \aa$ (resp. $N\bb$ in $2d$) grows with $N$. In the large $N$ limit the ground state energy per particle is thus described by a Gross-Pitaevskii energy functional where 
the gradient term  becomes negligible compared to the other terms, see \cite[Thms.6.3 and 6.6]{LSSY}.  
The Thomas-Fermi limit is particularly relevant to derive the effective behaviour of rapidly rotating Bose gases in anharmonic traps, see \cite{BCPY}. 
Recently, a simultaneous Thomas-Fermi and mean field scaling was considered in \cite{DG}. 

\bigskip \bigskip

{\bf Two dimensions.} 
Scaling limits interpolating between the mean field and Gross-Pitaevskii scalings (hence corresponding to those described by \eqref{eq:HN-beta} in the three dimensional case) are given in two dimensions by the family of Hamiltonians:
\be \label{eq:HN-beta2d}
 H_{N,2d}^{(\b)}= \sum_{i=1}^N -\D_{x_i} + \sum_{1\leq i < j \leq N} N^{2\b-1} V(N^\b (x_i- x_j)) \qquad \text{on} \qquad L^2_s([0,1]^2)
\ee
for any $\b\geq 0$ independent on $N$. Note in particular that the choice $\b \in [0,1/2)$ corresponds to high density regimes, while setting $\b>1/2$ we are describing low density regimes. Eventually the parameter $\b \geq 0$ might be let to depend on $N$, provided that 
\be \label{eq:betamax}
\lim_{N \to \io} (\log N^\b)/N=0\,,
\ee
so that \eqref{eq:HN-beta2d} does not include the Gross-Pitaevskii regime. The Hamiltonian \eqref{eq:HN-beta2d} describes a weakly interacting system with $\l \sim N^{-1}$, hence in particular, using the notation
$V_N(x)=N^{2\b-1} V(N^\b x)$, we have  $\bb \sim \widehat V_N(0)\sim N^{-1}$. This has to be compared with the Gross-Pitaevskii regime, where $\bb \sim N^{-1}$ is much smaller than the integral of the Gross-Pitaevskii potential, which is of order one.

Systems described by \eqref{eq:HN-beta2d} were first analysed in \cite{LNR,LNR1,NR}. The method used in these papers (see \cite{Rougerie} for a review) allows to obtain a proof of Bose-Einstein condensation and a derivation of the leading order term in the asymptotics for the ground state energy
(which is also valid for non positive potential) provided $\b<1$. These results were later extended in \cite{Caraci}, where an optimal bound for Bose-Einstein condensation has been obtained for all $\b>0$ (up to the threshold defined by \eqref{eq:betamax}), in the case of non negative potentials. The latter result is at the core of a proof of the validity of Bogoliubov theory for systems described by \eqref{eq:HN-beta2d}, see \cite{Caraci2}.  

\medskip
Regimes analogous to those described in \eqref{eq:HN-kappa}, less dilute than the Gross-Pitaevskii regime, can be obtained in two dimensions by considering  Hamiltonians of the form
\be \label{eq:HN-kappa2d}
H_{N,2d}^{(\kappa)}= \sum_{i=1}^N -\D_{x_i} + \sum_{1\leq i < j \leq N} e^{2N^{(1-\kappa)}} V( e^{N^{(1-\kappa)}} (x_i- x_j)) \qquad \text{on} \qquad L^2_s([0,1]^2)\,,
\ee
for $\kappa \in [0,1)$. Indeed with this choice of the interaction we have $\bb \sim N^{-1+\kappa}$. Note that regimes as in \eqref{eq:HN-kappa2d} interpolates between the Gross-Pitaevskii regime $\kappa=0$ and the thermodynamic limit (where the interaction is not rescaled and $\bb\sim 1$). From the point of view of length scales, the condition $\bb \sim N^{-1+\kappa}$ is equivalent to fixing
\be \label{eq:L2d}
L \sim (\r \bb)^{-1/2} \,\bb^{-\a}\,, \qquad \a = \kappa/2(1-\kappa)
\ee
A proof of BEC for regimes described by \eqref{eq:HN-kappa2d} for all $\kappa<1/6$ (corresponding to length scales as in \eqref{eq:L2d} with $\a<1/10$) can be obtained using the methods \cite{LSY2}. For weak interactions, order by orders bounds in perturbation theory compatible with the occurrence of condensation were proved in \cite{CG} up to length scales as in \eqref{eq:L2d} with $\a<1$ (corresponding to $\kappa<2/3$). An upper bound for the ground state energy compatible with \eqref{eq:LHY2d} and valid for all $\a>0$ (namely for all $\kappa<1$) has been recently shown in  \cite{FGJMO}.

\bigskip

{\bf Summarising.} In this section we reviewed recent results where the occurrence of condensation and the validity of Bogoliubov's prediction for the ground state energy were  proved for systems defined on finite length scales.
In the next two sections we will discuss how the progress in proving Bose-Einstein condensation and the validity of Bogoliubov theory on finite, but increasingly larger, boxes has been linked to 
the advances in the understanding of the behaviour of interacting bosons in the thermodynamic limit. The mathematical tools  allowing to obtain information about thermodynamic functions, starting from the properties of  finite-size systems, are usually referred to as {\it localization methods}. Since different localization methods are required to obtain lower and upper bounds compatible with \eqref{eq:LHY} and \eqref{eq:LHY2d}, we will discuss the two problems separately, trying to emphasise the major difficulties to be overcome in the two situations.

\section{Thermodynamic limit: lower bounds} \label{sec:LB}

The aim of this section is to review the methods allowing to reduce the problem of obtaining a lower bound on the specific ground state energy of \eqref{eq:HT}, to a proof of Bose-Einstein condensation on a sufficiently large box.
We begin with the pioneering paper by Lieb-Yngvason \cite{LY} back in 1998, to arrive to the very recent localization technique developed by Briezke, Fournais and Solovej \cite{BrFS}, finally allowing to capture
the second order correction in \eqref{eq:LHY} and \eqref{eq:LHY2d}, as a lower bound, for general non negative interactions, including the hard core case, both in three \cite{FS,FS2} and in two dimensions \cite{FGJMO}.

\medskip

{\bf The Neumann bracketing.} It may seem a long time ago, but it was only in 1998 
that a lower bound compatible with the leading order in \eqref{eq:LHY} was first obtained \cite{LY}.  
To achieve this result Lieb and Yngvason implemented a localization method usually referred to as {\it the box method} or {\it Neumann bracketing} (see \cite[Eq.(4)~--~ (12)]{LY}). Namely, they decomposed the system described by \eqref{eq:HT} on a box of side $T$ (with Neumann boundary conditions) into boxes of side $L$ with Neumann boundary conditions as well, getting the lower bound 
\be\label{eq:box_method}
    \frac{E_3(N,T)}{N} \geq \frac{1}{\r L^3} \inf_{c_n}\sum_{n\geq 0}c_n E_3(n,L)\,,
\ee
where the coefficients $c_n$ (relative number of boxes containing $n$ particles) have to satisfies the constraints $\sum_{n\geq 0}c_n=1$ and $\sum_{n\geq 0}nc_n=\r L^3.$ The difficulty then relies in proving that to minimize the total energy the number of particles has to be approximately uniform in all boxes. To do so the authors exploit the sub-additivity property of the ground state energy, which follows from the positivity of the interaction. 

Eq. \eqref{eq:box_method} reduces the problem of getting a lower bound for $E_3(N,T)$ to the study of the Bose gas in a box of side length  $L=\r^{-1/3}(\r \aa^3)^{-\d}$ for some small $\d>0$, with Neumann boundary conditions. Here the condition $L\gg \r^{-1/3}$, ensures that each box contains a large number of particles $n$ so that $n(n-1) \simeq n^2$. The strategy used in \cite{LY} to study the last problem is based on two by now very famous steps. The first one is a generalization of a lemma due to Dyson (see \cite{Dy}), allowing to replace the original interaction by a more regular potential living on a longer scale, at the expense of part of the kinetic energy (we denote by $\eps>0$ the fraction of kinetic energy left after this step). The second step consists in regarding the regularized interaction as a (non negative) perturbation of the remaining kinetic energy, using Temple's inequality. Here a restriction on $L$ related to the gap of the Laplacian comes into play. Indeed, for Temple's inequality to make sense the kinetic energy of the first excited state in the small box, proportional to $ \eps L^{-2}$, has to control the expectation of the interaction energy on the ground state of the kinetic energy, which is of order $\aa \r^2 L^3$ (see \cite[Eq. (28)]{LY}). This constrains $L$ to be so that $ L \ll L_{\text{T}}:=\eps^{1/5} (\r \aa)^{-1/2} (\r \aa^3)^{1/10}$.

As a matter of fact, more conditions are required in the course of the proof to control error terms, which are met for  
$L \sim \r^{-1/3} (\r \aa^3)^{-1/51} 
\ll L_T$. 

The same strategy also works in two dimensions \cite{LY2d}, provided a suitable modification of the Dyson Lemma, to account for the logarithmic behaviour of the two dimensional scattering function. 
The proof of the leading order term in \eqref{eq:LHY2d} is thus achieved by studying the problem on a box with side length of the order $ \r^{-1/2} \,|\log(\r \aa^2)|^{1/10}$.

The construction in \cite{LY, LY2d} allows also to show the occurrence of Bose-Einstein condensation in $3d$ and $2d$ over  lengths scale of order
$(\r \aa)^{-1/2}(\r\aa^3)^{-1/34}$
\cite{LS}, and  $\r^{-1/2} |\log(\r \aa^2)|^{1/10}$ \cite{LSY2} respectively (see the two paragraphs after \eqref{eq:L3d} and \eqref{eq:L2d} above for  a summary  about the length scales over which condensation can be currently shown).

\medskip

{\bf The sliding method.} For almost ten years  it was not clear how to go beyond the methods of \cite{LY} to rigorously derive the second term in \eqref{eq:LHY}. In \cite{GiuS} -- following a strategy applied first to the bosonic jellium in \cite{LS-jellium, LS-charged} -- Giuliani and Seiringer finally provided a rigorous derivation of \eqref{eq:LHY} in a weak coupling and high density regime satisfying \eqref{eq:Bog-exact} with $a_0/R = (\r \aa^3)^\g$ and $\g$ close to $1/2$. As previously discussed, under the conditions \eqref{eq:Bog-exact} Bogoliubov's heuristic strategy can be proven to be correct. In particular it is possible to show that in the ground state of \eqref{eq:HT} the ratio among the number $n_+$ of particles orthogonal to the condensate and the total number $n$ of particles in a box of side length $L$ is bounded by \cite[Lemma III.4]{GiuS} 
\[ \label{eq:BEC-GiuS}
\frac{n_+}{n} \leq C a_0 L^2 R^{-3}\,.
\]
In other words condensation occurs up to $L \ll (\r \aa)^{-1/2}(\r \aa^3)^{(1-3\g)/2}$. We remark that with this side length the usual localization error $L^{-2}$ would be of order $\r \aa (\r \aa)^{1/2 +3(\g-1/2)}$, which is larger than the Lee-Huang-Yang second order correction for any $\g<1/2$. In fact a key role in the paper by Giuliani and Seiringer is played by the sliding method introduced in \cite{CLY}, where a lower bound is achieved by decomposing the thermodynamic box into small boxes, and then averaging over all possible locations of the boxes.

Let us state the result of the sliding localization, as used in \cite{GiuS}. Let $\L_T=[-T/2,T/2]^3$ be a three dimensional torus, and let us consider on $L^2_s(\L_T^{N})$ the Hamiltonian
\be \label{eq:Hprime}
H'_{T,N} = - \sum_{i=1}^N \D_{x_i} + \frac{a_0}{R_0^3} \sum_{1\leq i < j \leq N} v_{R_0}(x_i-x_j)  - 4 \pi N \r a_0 \,,
\ee
with $v_{R_0}(x)= \sum_{n \in \bZ^3} e^{-|x+n T|/R_0} $.  Note that $H'_{T,N}$ is obtained by considering in \eqref{eq:HT} an interaction of the form $V(x)= a_0 R_0^{-3} v_{R_0}(x)$ and shifting the ground state by the leading order term in \eqref{eq:LHY}, after having used the Born series to replace $\aa$ with $a_0$, see \eqref{eq:Born}. Hence the ground state of $H'_{T,N}$ will be negative.

We now introduce a function localizing the particles on lengths smaller than $T$. We fix a small parameter $t=(\r a_0^3)^{\t}$, and define $\chi \in C_0^\io(\bR^3)$, spherically symmetric,  $0 \leq \chi \leq 1$, such that $\chi(x)$ is supported in  $[(-1+t)/2, (1-t)/2]^3$  and is identically one in the smaller box   $[(-1+2t)/2, (1-2t)/2]^3$. Then we denote by $\chi_L(x)$ the function on the torus $\L_T$ defined by $\chi_{L}(x)= \sum_{n \in \bZ^3} \chi(L^{-1}(x+n T)) $ and with
\[
w_{R_<}(x,y)= \chi(x/L) e^{-|x-y|/R_<}\chi(y/L)
\]
a localized interaction with range $R_<$. On $L^2([-L/2, L/2]^{3n})$ we define the small box Hamiltonian
\be \begin{split} \label{eq:HLn}
& H_{L,R}^n = - \sum_{j=1}^n \D_L^{(j)} \\
& + \frac{ a_0 I R }{R_0^4} \bigg[ \sum_{1\leq i < j \leq n} w_R(x_i,x_j) - \r \sum_{j=1}^n \int_{\bR^3}  w_R(x_j,y)dy +\frac{\r^2}{2} \int_{\bR^3\times \bR^3} w_R(x,y)dx dy\bigg]\,,
\end{split}\ee
where $\D_L^{(j)}$  is the Neumann Laplacian for the $j$-th particle in the cube $[-L/2,L/2]$ and  $I= (\int \chi^2(y) dy)^{-1}$. The following lemma holds:
\begin{lemma} Let $E'_{N,T}$ and $E_{n,L}^{(R)}$ be the ground state energy of the Hamiltonian defined in \eqref{eq:Hprime} and \eqref{eq:HLn} respectively. Then, for $t L/R_0$ large enough, there exists a function of the form $\o(t)= \text{const.} (t L / R_0)^{-1} $ such that if we set $R^{-1}= R_0^{-1} + \o(t)/L$ we have
\be \label{eq:GSloc}
\lim_{N \to \io} \frac{E'_{N,T}}{N}\geq  \frac{1}{\r L^3} \inf_n E_{n,L}^{(R)} - \frac{a_0 \o(t) R}{2L R_0^3}\,.
\ee
\end{lemma}
The last term on the r.h.s. of \eqref{eq:GSloc} is the error due to the localization to the smaller box. It is proportional to $ \frac{1}{L^2} \frac{a_0}{t R_0}$. Hence, w.r.t. the `standard' localization error $L^{-2}$
we gain some extra smallness from the high density condition $a_0/R_0 = (\r a_0^3)^{\g}$, for $\g \in (1/3, 1/2)$.
In particular, by choosing
\[
t = (\r a_0^3)^{\t}\,, \quad L= (\r a_0)^{-1/2} (\r a_0^3)^{-b}\,, \quad a_0/R_0 = (\r a_0^3)^{1/2 -\nu}
\]
we have that the localization error is smaller than the second-order correction to the ground state energy if $2b >\nu +\t$. Hence, the high density condition in \cite{GiuS} has two key advantages: first it makes the analysis on the small box `simpler' since one only needs to follow Bogoliubov's strategy; moreover it allows to improve the localization error. Note, however, that the conditions in \cite{GiuS} correspond to a very high density regime. In particular the authors fix $\nu <1/69$ (see Fig.~\ref{Fig:Bog}).

\bigskip

{\bf The small box localization.} 
Here we discuss the localization method applied in \cite{BrFS}  to obtain a lower bound which captures the correct second order with the wrong constant. The main difference w.r.t. \cite{GiuS} is that here the Hamiltonian in the small box exhibits a modified kinetic energy. Consider the Hamiltonian $H_T''$ on the bosonic Fock space $\mathcal{F}_S(L^2(\Lambda_T))$ built on the box $\Lambda_T=[-T/2,T/2]^3$ with Dirichlet boundary conditions and whose restriction on the $N$-particle subspace is given by 
\begin{equation*}\label{eq:Hprimeprime}
    H_{T,N}''=-\sum_{i=1}^N\Delta_{x_i}+\sum_{i<j}v(x_i-x_j)-8\pi N \rho \aa\,.
\end{equation*}
Note that the Hamiltonian $H_{T,N}''$ is formulated in the grand canonical setting and the term $-8 \pi N \r \aa $ in the above definition can be interpreted as the addition of a term $-\m N$ fixing the total number of particles to $N$, with  $\mu= 8 \pi \r \aa $ the expected chemical potential on the basis of Bogoliubov theory. 
Clearly, an estimate from below on $H_{T,N}''$ provides a lower bound  on the specific ground state energy $e_3(\rho)$ defined in \eqref{eq:e_d} since the canonical ground state can be considered as a trial state for the grand canonical Hamiltonian. More precisely,
\[
    e_3(\rho)\geq \lim_{|\Lambda_T|\to\infty}\frac{1}{|\Lambda_T|}\inf_{\psi\in \mathcal{F}_S(L^2(\Lambda_T))}\frac{\langle\psi,H_{T,N}''\psi\rangle}{\|\psi\|^2}+8\pi\aa\rho^2\,.
\]
We are going to localize on a small box of side length $L=K( \rho \aa)^{-1/2}$ for $K>0$ to be chosen sufficiently small but independent of $\rho.$ To write the Hamiltonian in the small box we first introduce the localization function $\chi\in C^\infty_0(\mathbb{R}^3)$ even s.t. $0\leq \chi\leq 1$, $\int \chi^2=1$ and $\supp\chi\subset[-1/2,1/2]^   3.$ For a given $u\in \mathbb{R}^3$ we set  $\chi_u(x)=\chi(x/L-u),$ then the potential in the small box $uL+[-L/u,L/u]^3$ centered in $uL$ is given by
\[
    w_u(x,y)=\chi_u(x)\frac{v(x-y)}{(\chi\ast\chi)((x-y)/L)}\chi_u(y)\,.
\]
We also introduce the notation 
\[
    \mathcal{W}_u(x_1,\dots,x_N)=\sum_{i<j}w_u(x_i,x_j)+\sum_{i=1}^N-\rho\int w_u(x_i,y)f(x_i-y)dy\,,
\]
where $f$ is the scattering solution defined in \eqref{eq:scatt_3d}. A direct calculation shows that
\begin{equation}\label{eq:pot}
    \sum_{i<j}v(x_i-x_j)-8\pi N\rho\aa= \int_{L^{-1}(\Lambda_T+B(0,L))}\mathcal{W}_u(x_1,\dots,x_N)du\,.
\end{equation}
On the other hand, the usual kinetic energy is replaced in the small box by the operator
\[
    \mathcal{T}_u=Q_u\Big[\chi_u\big(-\Delta-\frac14(sL)^{-2}\big)_+\chi_u+b/L^2\Big]Q_u\,,
\]
where $s,b>0$, $(\cdot)_+$ denotes the positive part and $Q_u$ is the orthogonal projection defined by $Q_u\phi=\theta_u\phi-L^{-3}(\theta_u,\phi)\theta_u$ with $\theta_u$ the indicator function of the box centered in $uL.$ Let us comment on the operator $\mathcal{T}_u$. First we note that it vanishes on constant functions, moreover the last term in the square bracket plays the role of a Neumann gap, namely $\cT_u$ on functions orthogonal to constants is bounded below by at least $bL^{-2}.$ Finally, $(-\Delta- \frac14(sL)^{-2})_+$ is to be understood as the multiplication operator  $(p^2-\frac14(sL)^{-2})_+$ in Fourier space,
hence momenta $p$ of order $(\rho\aa)^{1/2},$ which are relevant to capture the second order in the energy (as discussed in Sec. \ref{sec:intro}) are correctly recovered.\\
One can show that the kinetic energy in the thermodynamic box can be bounded by an integral over the kinetic operators $\mathcal{T}_{u}$ on the small boxes, namely there exists a $b$ such that, if $s$ is small enough then
\begin{equation}\label{eq:kin}
    -\Delta\geq \int_{L^{-1}(\Lambda_T+B(0,L))}\mathcal{T}_{u} \,du\,.
\end{equation}
The proof of the above inequality relies on the general fact that given $\mathcal{K}:\mathbb{R}^3\to[0,\infty)$ symmetric and polynomially bounded, then the operator 
\[
    T=\int_{\mathbb{R}^3}Q_u\chi_u\mathcal{K}(-iL\nabla)\chi_uQ_u \,du
\]
is such that $T=F(-iL\nabla)$, \ie in Fourier space it is the multiplication operator times $F(p)$ where
\[
    F(p)=(2\pi)^{-3}\mathcal{K}\ast|\widehat{\chi}|^2(p)-2(2\pi)^{-3}\widehat{\theta}(p)\widehat{\chi}\ast(\mathcal{K}\widehat{\chi})(p)+(2\pi)^{-3}\Big(\int \mathcal{K}|\widehat{\chi}|^2\Big)\widehat{\theta}(p)^2\,,
\]
where we recall the definition \eqref{eq:Fourier} of Fourier transform $\widehat{f}(p)$ of $f$.

While the proof of \eqref{eq:kin} is quite technical, it can be regarded as an extension of the IMS localization formula. Indeed, the latter can be stated as follows: the operator $\int_{\mathbb{R}^3} \chi_u\mathcal{K}(-i\nabla)\chi_u$ with $\mathcal{K}(p)=p^2$ acts in Fourier space as the multiplication by
\[
  (2\pi)^{-3}\mathcal{K}\ast|\widehat{\chi}|^2=p^2+\int |\nabla\chi|^2\,.
\]

We define the Hamiltonian $H_{L,u}$ in Fock space acting on the $N$ particle sector as
\[
    H_{L,u,N}=\sum_{i=1}^N\mathcal{T}_{u,i}+\mathcal{W}_u(x_1,\dots,x_N)\,,
\]
and noting that $H_{L,u}$ and $H_{L,u'}$ are unitarily equivalent one concludes from \eqref{eq:pot}, \eqref{eq:kin}
\begin{equation*}\label{eq:bound_small_box}
   \lim_{|\Lambda_T|\to\infty}\frac{1}{|\Lambda_T|}\inf_{\psi\in \mathcal{F}_s(L^2(\Lambda_T))}\frac{\langle \psi,H_{T,N}''\psi\rangle}{\|\psi\|^2}\geq\frac1{ L^{3}}\inf\mathrm{spec}H_{L,0}\,.
\end{equation*}
The main result in \cite{BrFS} is the following estimate on the Hamiltonian in the small box:
\begin{equation*}\label{eq:en_small_box}
    H_{L,0}\geq -4\pi \aa\rho^2L^3-C\rho^2\aa\sqrt{\rho \aa^3}L^3-C\rho^3\aa^2R^2L^3\,,
\end{equation*}
where $R$ is the range of the interaction. Hence, assuming $\rho$ to be small enough
\[
    e_3(\rho)\geq 4\pi\aa\r^2\big(1-C(\sqrt{\r\aa^3}+R^2\aa\r)\big)\,.
\]  
Note that the localization method described above allows to prove BEC on length scales $L=C_L(\r\aa)^{-1/2} (\r\aa^3)^{-\a}$ for $\a < 1/4$  as shown in \cite{F}.

\bigskip

{\bf The double box localization.} 
This is the method introduced in \cite{BriS} to extend the result obtained in \cite{GiuS} to  scaled potentials of the form
\[
    v_R(x)=\frac1{R^3}v(x/R)
\]
for a given $v$ non-negative, spherically symmetric and compactly supported under the assumptions (see \eg \cite[Thm 1.1]{BriS}) that
\be\label{eq:scaling_BFS}
    \lim_{\r\to 0}\frac{R}{\aa}(\r\aa^3)^{1/2}=0, \qquad \text{ and }\qquad \lim_{\r\to 0}R\r^{1/3}(\r\aa^3)^{-1/30}=\infty\,.
\ee
In particular, assuming $\aa/R=(\r\aa^3)^\gamma$, the condition \eqref{eq:scaling_BFS} implies $3/10<\g<1/2.$ Hence, in contrast to \cite{GiuS} where $1/2-1/69<\gamma<1/2$ was needed, here also {\it low density} regimes are handed. In fact, choosing $3/10<\gamma<1/3$, we get $R\r^{1/3}=(\r\aa^3)^{1/3-\g}\ll1$ (see Fig. \ref{Fig:Bog}). 

The novelty in the localization used in \cite{BriS} is to proceed in two steps:  the system is first localized to ``large" boxes of side length $L\gg(\r\aa)^{-1/2}$ and then to ``small'' boxes smaller than $(\r\aa)^{-1/2}$. 
The method, which is an ingenious extension of the sliding method described above, comes with a subtle modification of the kinetic energy, see \cite[Eq.(57)]{BriS} for the  large box kinetic operator, and \cite[Eq.(64)]{BriS} for the small box one.  On the small box, slightly smaller than the Gross-Pitaevskii scale: i) the number of excitations is much smaller than the total number of particles; ii) due to conditions \eqref{eq:scaling_BFS}, which imply the chain of inequalities in \eqref{eq:Bog-exact}, the heuristics proposed by Bogoliubov can be proven to be correct; in particular error terms can be absorbed in the gap.
In a second stage, the estimates on the energy and number of excitation obtained on the small boxes yield  a priori bounds on the energy, number of particles (which will be shown to be close to $\r L^3$) and excited particles in the large box  
\cite[Sec. 6]{BriS}.  

The localization technique from \cite{BriS} was exploited by Fournais and Solovej in the papers \cite{FS, FS2} allowing to get a lower bound compatible with \eqref{eq:LHY}, for general non negative interactions, thus closing a longstanding problem in mathematical physics. This result was achieved thanks to a more refined analysis of the large box Hamiltonian, allowing the authors to extract the missing contributions to the ground state energy hidden in the cubic and quartic terms neglected in Bogoliubov theory, and which could not be extracted by restricting to quasi-free states as in \cite{NRS1,NRS2}. We refer to \cite[Sec. 2]{FS2} for a description of the general strategy pursued in \cite{FS,FS2} and the additional complications to be overcome to treat potentials with large $L^1$ norm.

The same localization method is at the core of the recent paper \cite{FGJMO}, where a lower (and upper) bound matching \eqref{eq:LHY2d} has been recently obtained. There, similarly as in three dimensions, the small and large boxes are taken to be slightly smaller or larger than the Gross-Pitaevskii scale $(\r \bb)^{-1/2}$. In two dimensions, however, the authors had to face similar challanges as in the $3d$ hard core case, even for regular potentials, being the effective parameter $\bb$ much smaller than the integral of the potential.

\bigskip

{\bf Summarizing.} 
A major challenge for obtaining a second order lower bound compatible with \eqref{eq:LHY} has been the development of a localization method allowing to obtain a Neumann gap both on the small and large boxes (to be able to absorb error terms), still keeping essentially the original kinetic energy at the relevant Bogolubov scales (see \cite[Remark 5.5]{FS} for a discussion of the localized kinetic energy in \cite{BriS, FS}). Thanks to this localization bound, the problem is reduced to showing the  
validity of Bogoliubov theory on a box of side length slightly larger than the Gross-Pitaevskii length scale.  On this scale, it is essential to extract some missing energy from cubic and quartic terms neglected in Bogoliubov theory, which cannot be captured by restricting to quasi-free states. Indeed one needs to take into account correlations among triples of excitations. This idea, which appeared first in the trial state by Yau-Yin \cite{YY}, is also at the core of the works \cite{BBCS4} and \cite{FS} (where it appeared formulated with very different languages though), and has been later exploited in all works aimed at applying Bogoliubov's approach at the Gross-Pitaevskii scale and beyond (see references in Sec. \ref{sec:scalings}.)

An alternative route to the double box localization would require studying directly the Neumann problem rather than the periodic one, on a sufficiently large box. 
Very recently, Boccato and Seiringer \cite{BoSei} managed to study the properties of three dimensional Bose gases with Neumann boundary conditions directly on boxes of side length $L \sim (\r \aa)^{-1/2}$. Via Neumann bracketing this implies a
lower bound for the leading order of the ground state energy per particle of a Bose gas in the thermodynamic limit, with almost optimal error (the error in \cite{BoSei} is $\r \aa (\r \aa^3)^{1/2}|\log(\r \aa^3)|$ due to a finite size effect related to the Neumann boundary conditions, see \cite[Eq.(1.12)]{BoSei}). 
The extension of this result to 
boxes of side length larger than $(\r \aa)^{-1/2}$ would allow for an alternative proof of a lower bound compatible with \eqref{eq:LHY}, 
without need of refined localization techniques.


\section{Thermodynamic limit: upper bounds} \label{sec:UB}

In this section we review the methods developed in the literature to get an upper bound on the specific ground state energy $e_d(\rho)$ defined in \eqref{eq:e_d}. 

An upper bound correct at leading order was first obtained in \cite{Dy} for the three dimensional hard-core Bose gas and later extended in \cite{LSY}  to cover all non-negative interactions and in \cite{LY2d} to treat the two dimensional case. To get an upper bound on the ground state energy it is enough to evaluate the energy on an appropriate state. In the aforementioned papers the authors considered a trial state of the form

\be\label{eq:dyson}
    \Psi(x_1,\dots,x_N)=F_1( x_1)F_2( x_2)\dots F_N(x_1,\dots,x_N)
\ee
where 
\[
    F_i( x_1,\dots,x_i)= f_\ell(t_i),\quad t_i=\min_{j=1,\dots,i-1}|x_i-x_j|\,,
\]
and the function $f_\ell$ is a modified version (normalized to 1 at length $\ell$) of the scattering function $f$  defined in \eqref{eq:scatt_3d}  in three dimensions, and it is defined by \eqref{eq:scatt_2d} in two dimensions. In \eqref{eq:dyson} the cutoff is put at a length scale of the order $\r^{-1/d}$ (see \cite[Eq. (2.19)]{LSSY} for the precise definition).
The state \eqref{eq:dyson} has to be thought of as a modification of the so-called Jastrow factor, namely the wave function 
\be \label{eq:jastrow}
    \prod_{i<j}f_\ell(x_i-x_j)
\ee
considered in \cite{J}. The idea behind the Jastrow trial state \eqref{eq:jastrow} is the following: in contrast with the non-interacting case where the ground state was just the constant wave function, in the interacting case, correlations have to be considered when two particles come closer. In particular, in the dilute regime, it seems reasonable to expect that only two-body correlations matter, and that two-body correlations are well described by the zero energy scattering function.

 The trial state \eqref{eq:jastrow} is relatively manageable  for $\ell \ll (\r \aa)^{-1/2}$ in $3d$ and  $\ell \ll (\r \bb)^{-1/2}$ in $2d$; we refer the reader to \cite[Sect. 2.1]{LSSY} and \cite[Sect. 2]{BCOPS-rev} for a computation of the energy of \eqref{eq:dyson} and \eqref{eq:jastrow} respectively. However, it seems difficult to compute the energy of a state as in \eqref{eq:jastrow} for $\ell$ larger than the healing length, a choice which would be needed to capture the second order term in \eqref{eq:LHY}, as discussed in Sec.\ref{sec:intro}.
In fact, the next order in the energy was first tackled in three dimensions in \cite{ESY} with a different method, closer to Bogoliubov's heuristics, as described in Sec.\ref{sec:intro}. Finally, it was in \cite{YY} that the Lee-Huang-Yang correction was obtained as an upper bound, by extending the trial state considered in \cite{ESY}, including correlation among triples of excited particles ({\it soft-pair creation} in the language of the paper) and exploiting a localization method discussed in the next subsection.  The approach of \cite{YY} has been reviewed and adapted to a grand canonical setting in \cite{Aaen}. The same localization technique, with a simpler trial state, was then used in \cite{BCS} to cover all potentials  $V \in L^3.$ Note that a second order upper bound also covering the hard sphere case in three dimensions is still missing. 

On the other hand,  in the two dimensional case an upper bound in agreement with \eqref{eq:LHY2d} was recently obtained in \cite{FGJMO} for a very general class of potentials including hard-core interactions. This result was achieved by combining the localization method of \cite{YY} (applied to the  two dimensional case) with a trial state obtained multiplying the Jastrow function by a quasi-free state. Note that a similar trial state has been used in \cite{BCOPS} to get a second-order upper bound on the energy for hard core bosons in the three-dimensional Gross-Pitaevskii regime. The reason why in three dimension the same idea cannot be applied to the thermodynamic limit will be discussed at the end of this section.


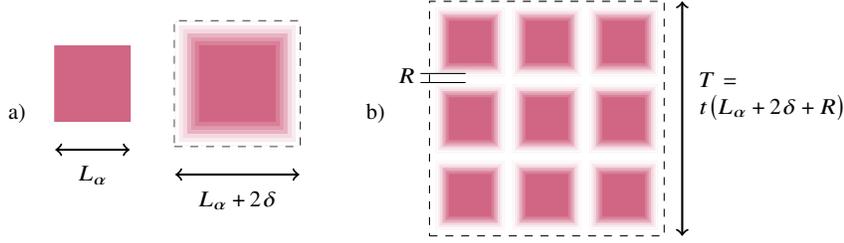
\begin{figure}[t]
\centering   a) \hskip 0.3cm 
\begin{tikzpicture}[scale =1.1,
    every path/.style = {},
 ]
 \begin{scope}
\draw[thick, color=myred!60!white, fill=myred!60!white] (0.05,0.05) rectangle (0.95,0.95);
 \draw[color=black,<->, thick] (0.05,-0.3) -- (0.95,-0.3);
 \node[below, color=black] at (0.5,-0.4){$L_\a$};
\end{scope}
\end{tikzpicture}  \hskip 0.5cm
\begin{tikzpicture}[scale =1.1,
    every path/.style = {},
 ]
 \begin{scope}
 \draw[thick, color=myred!60!white, fill=myred!60!white] (0.05,0.05) rectangle (0.95,0.95);
 \draw[dashed, color=black] (-0.26,-0.26) rectangle (1.26,1.26);
 \draw[color=myred!5!white] (-0.25,-0.25) rectangle (1.25,1.25);
\draw[color=myred!8!white, fill=myred!8!white] (-0.2,-0.2) rectangle (1.2,1.2);
\draw[color=myred!15!white, fill=myred!15!white] (-0.15,-0.15) rectangle (1.15,1.15);
\draw[color=myred!30!white, fill=myred!30!white] (-0.1,-0.1) rectangle (1.1,1.1);
\draw[color=myred!40!white, fill=myred!40!white] (-0.05,-0.05) rectangle (1.05,1.05);
\draw[thick, color=myred!50!white, fill=myred!50!white] (0,0) rectangle (1,1);
\draw[thick, color=myred!60!white, fill=myred!60!white] (0.05,0.05) rectangle (0.95,0.95);
\draw[color=black,<->, thick] (-0.25,-0.6) -- (1.25,-0.6);
\node[below, color=black] at (0.5,-0.7){$L_\a + 2 \d$};
\end{scope}
\end{tikzpicture} \hskip 0.8cm  b)
\begin{tikzpicture}[scale =0.6,
    every path/.style = {},
 ]
 \begin{scope}
 \draw[thick, color=myred!60!white, fill=myred!60!white] (0.05,0.05) rectangle (0.95,0.95);
 \draw[color=myred!5!white] (-0.25,-0.25) rectangle (1.25,1.25);
\draw[color=myred!8!white, fill=myred!8!white] (-0.2,-0.2) rectangle (1.2,1.2);
\draw[color=myred!15!white, fill=myred!15!white] (-0.15,-0.15) rectangle (1.15,1.15);
\draw[color=myred!30!white, fill=myred!30!white] (-0.1,-0.1) rectangle (1.1,1.1);
\draw[color=myred!40!white, fill=myred!40!white] (-0.05,-0.05) rectangle (1.05,1.05);
\draw[thick, color=myred!50!white, fill=myred!50!white] (0,0) rectangle (1,1);
\draw[thick, color=myred!60!white, fill=myred!60!white] (0.05,0.05) rectangle (0.95,0.95);
\draw[color=black](-0.6, -0.2)--(0.4, -0.2);
\draw[color=black](-0.6, -0.4)--(0.4, -0.4);
\node[above, color=black] at (-0.9,-0.6) {\footnotesize $R$};
\draw[color=black, dashed]  (-0.4,1.4) rectangle (4.8,-3.8);
\draw[color=black,<->, thick] (5.2,1.4) -- (5.2,-3.8);
\node[below, color=black] at (6,0) { $T=$};
\node[below, color=black] at (7.2,-0.5) { $t \big(L_\a + 2 \d  + R \big)$};
 \draw[color=myred!5!white] (1.45,-0.25) rectangle (2.95,1.25);
\draw[color=myred!8!white, fill=myred!8!white] (1.5,-0.2) rectangle (2.9,1.2);
\draw[color=myred!15!white, fill=myred!15!white] (1.55,-0.15) rectangle (2.85,1.15);
\draw[color=myred!30!white, fill=myred!30!white] (1.6,-0.1) rectangle (2.8,1.1);
\draw[color=myred!40!white, fill=myred!40!white] (1.65,-0.05) rectangle (2.75,1.05);
\draw[thick, color=myred!50!white, fill=myred!50!white] (1.7,0) rectangle (2.7,1);
\draw[thick, color=myred!60!white, fill=myred!60!white] (1.75,0.05) rectangle (2.65,0.95);
\draw[color=myred!5!white] (3.15,-0.25) rectangle (4.65,1.25);
\draw[color=myred!8!white, fill=myred!8!white] (3.2,-0.2) rectangle (4.6,1.2);
\draw[color=myred!15!white, fill=myred!15!white] (3.25,-0.15) rectangle (4.55,1.15);
\draw[color=myred!30!white, fill=myred!30!white] (3.3,-0.1) rectangle (4.50,1.1);
\draw[color=myred!40!white, fill=myred!40!white] (3.35,-0.05) rectangle (4.45,1.05);
\draw[thick, color=myred!50!white, fill=myred!50!white] (3.4,0) rectangle (4.4,1);
\draw[thick, color=myred!60!white, fill=myred!60!white] (3.45,0.05) rectangle (4.35,0.95);
 \draw[color=myred!5!white] (-0.25,-1.95) rectangle (1.25,-0.45);
\draw[color=myred!8!white, fill=myred!8!white] (-0.2,-1.9) rectangle (1.2,-0.5);
\draw[color=myred!15!white, fill=myred!15!white] (-0.15,-1.85) rectangle (1.15,-0.55);
\draw[color=myred!30!white, fill=myred!30!white] (-0.1,-1.8) rectangle (1.1,-0.6);
\draw[color=myred!40!white, fill=myred!40!white] (-0.05,-1.75) rectangle (1.05,-0.65);
\draw[thick, color=myred!50!white, fill=myred!50!white] (0,-1.7) rectangle (1,-0.7);
\draw[thick, color=myred!60!white, fill=myred!60!white] (0.05,-1.65) rectangle (0.95,-0.75);
 \draw[color=myred!5!white] (1.45,-1.95) rectangle (2.95,-0.45);
\draw[color=myred!8!white, fill=myred!8!white] (1.5,-1.9) rectangle (2.9,-0.5);
\draw[color=myred!15!white, fill=myred!15!white] (1.55,-1.85) rectangle (2.85,-0.55);
\draw[color=myred!30!white, fill=myred!30!white] (1.6,-1.8) rectangle (2.8,-0.6);
\draw[color=myred!40!white, fill=myred!40!white] (1.65,-1.75) rectangle (2.75,-0.65);
\draw[thick, color=myred!50!white, fill=myred!50!white] (1.7,-1.7) rectangle (2.7,-0.7);
\draw[thick, color=myred!60!white, fill=myred!60!white] (1.75,-1.65) rectangle (2.65,-0.75);
\draw[color=myred!5!white] (3.15,-1.95) rectangle (4.65,-0.45);
\draw[color=myred!8!white, fill=myred!8!white] (3.2,-1.9) rectangle (4.6,-0.5);
\draw[color=myred!15!white, fill=myred!15!white] (3.25,-1.85) rectangle (4.55,-0.55);
\draw[color=myred!30!white, fill=myred!30!white] (3.3,-1.8) rectangle (4.5,-0.6);
\draw[color=myred!40!white, fill=myred!40!white] (3.35,-1.75) rectangle (4.45,-0.65);
\draw[thick, color=myred!50!white, fill=myred!50!white] (3.4,-1.7) rectangle (4.4,-0.7);
\draw[thick, color=myred!60!white, fill=myred!60!white] (3.45,-1.65) rectangle (4.35,-0.75);
\draw[color=myred!5!white] (-0.25,-3.65) rectangle (1.25,-2.15);
\draw[color=myred!8!white, fill=myred!8!white] (-0.2,-3.6) rectangle (1.2,-2.2);
\draw[color=myred!15!white, fill=myred!15!white] (-0.15,-3.55) rectangle (1.15,-2.25);
\draw[color=myred!30!white, fill=myred!30!white] (-0.1,-3.5) rectangle (1.1,-2.3);
\draw[color=myred!40!white, fill=myred!40!white] (-0.05,-3.45) rectangle (1.05,-2.35);
\draw[thick, color=myred!50!white, fill=myred!50!white] (0,-3.4) rectangle (1,-2.4);
\draw[thick, color=myred!60!white, fill=myred!60!white] (0.05,-3.35) rectangle (0.95,-2.45);
\draw[color=myred!5!white] (1.45,-3.65) rectangle (2.95,-2.15);
\draw[color=myred!8!white, fill=myred!8!white] (1.5,-3.6) rectangle (2.9,-2.2);
\draw[color=myred!15!white, fill=myred!15!white] (1.55,-3.55) rectangle (2.85,-2.25);
\draw[color=myred!30!white, fill=myred!30!white] (1.6,-3.5) rectangle (2.8,-2.3);
\draw[color=myred!40!white, fill=myred!40!white] (1.65,-3.45) rectangle (2.75,-2.35);
\draw[thick, color=myred!50!white, fill=myred!50!white] (1.7,-3.4) rectangle (2.7,-2.4);
\draw[thick, color=myred!60!white, fill=myred!60!white] (1.75,-3.35) rectangle (2.65,-2.45);
\draw[color=myred!5!white] (3.15,-3.65) rectangle (4.65,-2.15);
\draw[color=myred!8!white, fill=myred!8!white] (3.2,-3.6) rectangle (4.6,-2.2);
\draw[color=myred!15!white, fill=myred!15!white] (3.25,-3.55) rectangle (4.55,-2.25);
\draw[color=myred!30!white, fill=myred!30!white] (3.3,-3.5) rectangle (4.5,-2.3);
\draw[color=myred!40!white, fill=myred!40!white] (3.35,-3.45) rectangle (4.45,-2.35);
\draw[thick, color=myred!50!white, fill=myred!50!white] (3.4,-3.4) rectangle (4.4,-2.4);
\draw[thick, color=myred!60!white, fill=myred!60!white] (3.45,-3.35) rectangle (4.35,-2.45);
\end{scope}
\end{tikzpicture}
\caption{Schematic representation of the localization for the upper bound exploited in \cite{YY, Aaen, BCS, FGJMO}. The trial state on the thermodynamic box $T$  is obtained by gluing together $t^3$  small boxes with Dirichlet boundary conditions, leaving corridors of width $R$ among different boxes, see Fig. b). Each small box is obtained by enlarging a bit a box with side length $L_\a$ and periodic boundary condition, to obtain a trial state satisfying Dirichlet boundary conditions, see Fig. a). While on each small box the side length and the density are related, the thermodynamic limit is then taken by sending $t\to \io$ while keeping the density of the small boxes constant.  } 
\label{Fig:localization}

\end{figure}


\bigskip

{\bf Localization for second order upper bounds.} The proofs of the upper bounds matching \eqref{eq:LHY} (resp. \eqref{eq:LHY2d}) at second order obtained in \cite{YY, Aaen, BCS} (resp. \cite{FGJMO}) are all based on the construction of appropriate trial states for the Hamiltonian \eqref{eq:HT}. However, none of these trial states are constructed directly in $L^2_s (\L_T^N)$. Instead, it turns convenient (for reasons which will be clearer in a while) to build a trial state on smaller boxes, with side length $L_\a$ depending on $\r$, as in \eqref{eq:L3d} and \eqref{eq:L2d}. 

It is an easy observation that a thermodynamic trial state can be then built by gluing together trial states obtained on smaller boxes, if we consider Dirichlet rather than periodic boundary conditions, and if we impose that particles in different boxes do not interact (by forcing them to be at distance larger than the range of the potential), see Fig.\ref{Fig:localization}.  The core of the localization for the upper bound, whose standard proof \cite{R, YY, Aaen} is discussed in detail in \cite[App.~A]{BCS}  and generalized to non compact potentials in \cite[App.~A]{FGJMO}, is identifying the correct length scale for which the error coming from working in with periodic rather than Dirichlet boundary conditions is of smaller order with respect to the Lee-Huang-Yang second order correction.

In the following we will keep the discussion at the level of ideas, while we refer to \cite[Sec. 2.1]{YY} and  \cite[Prop.1.2]{BCS}, \cite[App. A]{FGJMO} for a precise statement of the localization result, in the canonical and in the grand canonical setting respectively.  To this end, it is sufficient to compare the leading order contribution to the energy in the case of Dirichlet and periodic boundary conditions. This is indeed the only source of error coming from the localization method described above.

\bigskip

{\bf Three dimensional case.} Let us consider $N$ bosons in the three dimensional torus $\Lambda_L$
with density $\r$. Then the ground state energy of the system satisfies \cite[Thm. 2.1]{LSSY}
\begin{equation*}\label{eq:Eperiodic}
E_3(N,L) = 4 \pi \aa N \rho \big(1 + o(1) \big)
\end{equation*}
for $\r \aa^3\ll1$. On the other hand, the energy of $N$ bosons in a box of side length $L$ with Dirichlet boundary condition is given, at leading order, by
\begin{equation*}\label{eq:EDirichlet}
	E_3^{(D)}(N,L) \simeq  N \min_{\ph \in \cC^\infty_c (\Lambda_L)} \Big[ \int |\nabla\ph|^2 + 4\pi \aa N \int |\ph|^4\Big] \,.
\end{equation*}
We now require that  
\begin{equation}\label{eq:difference}
\begin{split}
\frac{E_{3}^{(D)}(N,L)-E_3(N,L)}{N\rho} &\simeq  L^3 \min_{\ph \in \cC^\infty_c (\Lambda_L)} \Big[ \frac 1N \int |\nabla\ph|^2 + 4\pi \aa \int |\ph|^4\Big] - 4\pi\aa \\[0.2cm]
&\ll  C \aa (\rho \aa^3)^{1/2}\,.
\end{split}
\end{equation}
Even though we do not know the explicit expression of the minimizer on the first line of \eqref{eq:difference}, we know that it will be constant far from the boundary hence we can approximate it as
\begin{equation*}
	\label{eq:phapprox}
	\ph(x) = \begin{cases} \;\tfrac1{(L- 2\d)^{3/2}}\,, \quad & d(x, \dpr \L_L) < \d\,, \\
	\;	0\,, & \text{otherwise}\,.\end{cases}   
\end{equation*}  
Then we have
\[
\begin{split}
	\int |\ph|^4 &\sim \frac{1}{(L - 2\d)^6}(L-2\d)^3 =   \frac{1}{(L - 2\d)^3}\sim \frac{1}{L^3} \Big(1+ o\big(\d / L \big) \Big)\\
	\int |\nabla\ph|^2 & \sim \frac{1}{\d^2(L-2\d)^3}[L^3 - (L-2\d)^3] \sim \frac{1}{\d L}\Big(1 + o\big(\d/L\big)\Big)\,.
\end{split}\]
Indeed, being $N =\rho L^3$ and substituting in the expression in Eq. \eqref{eq:difference} we conclude
\be \label{eq:Dir-Per}
\begin{split}
\frac{E_3^{D}(N,L)}{N\rho} - \frac{E_3(N,L)}{N\rho} 
\simeq  \frac{1}{L} \Big(\aa \d + \frac{1}{\rho\d}\Big)\,.
\end{split}\ee
Optimization over $\d$ leads to the choice $\d = (\rho \aa)^{-1/2}$. Hence the condition in Eq. \eqref{eq:difference} is satisfied when 
\be \label{eq:localiz-UB}
L \gg (\rho \aa)^{-1/2} (\r \aa^3)^{-1/2}\,.
\ee
In particular, for $L_\a =(\rho \aa)^{-1/2} (\r \aa^3)^{-\a}$ being defined as in \eqref{eq:L3d}, this implies $\a > 1$ (or equivalently to study Hamiltonian of the form \eqref{eq:HN-kappa}, for $\kappa >1/2$).  

\medskip

{\it Remark.} Note that if one is interested in computing the leading order in the energy on the r.h.s. of \eqref{eq:LHY}, starting from a system were we imposed Dirichlet boundary conditions, the requirement $E_3^{(D)}(N,L) - E_3(N,L) \ll C N \r \aa$, together with \eqref{eq:Dir-Per}, leads to the condition $L \gg (\r \aa)^{-1/2}$. This confirms, as mentioned in the introduction, that the energy is spoiled at leading order if one localizes particles at length scales smaller than the healing length.

\bigskip

To conclude, let us comment on the requirement $\kappa>1/2$ corresponding to the condition \eqref{eq:localiz-UB}. One of the reasons why providing second order upper bounds for the ground state energy of Hamiltonians $\eqref{eq:HN-kappa}$ becomes increasingly difficult for increasing values of $\kappa$ is that the number of excitations needed to obtain the correct energy is of the order $N^{3\kappa/2}$ (see \eqref{eq:depletion}). In particular it is independent of $N$ in the  Gross-Pitaevskii regime, and then increases up to order $N$ (for $\kappa=2/3$). Moreover, the number of triples of excitations which have to be put on top of the quasi-free state and the condensate, to reproduce the Lee-Huang-Yang formula is of the order $N^{9\kappa/2 -2}$, see for example \cite[Prop. 2.2]{BCS}. While for $\kappa< 4/9$ this number goes to zero with $N$ (and one can control its effect by means of Gronwall type estimates, as done in the Gross-Pitaevskii regime, see \eg \cite[Prop. 4.2]{BBCS4}), for $\kappa>4/9$ the control of cubic excitations requires substantially new ideas.

\bigskip

{\bf Two-dimensional case.} As before, we consider $N$ bosons on the two dimensional torus $\L_L$ and we compare their ground state energy, denoted by $E_2(N,L)$ with the ground state energy of the same system in a box of side length $L$ and Dirichlet boundary condition, below denoted by $E_2^{(D)}(N,L)$. Reasoning as above we require
\[ \begin{split}
	\frac{E_2^{(D)}(N,L)}{N\rho} - \frac{E_2(N,L)}{N\rho}  \simeq \; & L^2 \min \Big\{ \frac 1 N \int |\nabla \ph|^2 + 4 \pi \bb \int |\ph|^4 \Big\} - 4 \pi \bb  \ll C \,\bb^2  \;,
\end{split}\]
being $N \r \bb^2$ the order at which we want to match \eqref{eq:LHY2d}. As in the $3d$ case we approximate the minimizer $\ph$ by
\[
\ph(x) = \begin{cases} \;1/(L- 2\d), \quad & d(x, \dpr \L_L) < \d\,, \\
	\; 0\,, & \text{otherwise}\,.\end{cases}  \,
\]
Proceeding as before we get
\be \begin{split} \label{eq:difference2d}
	\frac{E_2^{(D)}(N,L)}{N\rho} - \frac{E_2(N,L)}{N\rho}  \sim\; \frac 1 L \Big( \bb \d + \frac 1 {\r \d} \Big)  \sim\frac 1 L \sqrt{\bb/\rho} \ll \bb^2  \,,
\end{split}\ee
where in the last step we optimized over the choice of $\d$ by setting $\d^2 = 1/(\bb\rho)$. Then \eqref{eq:difference2d} is satisfied if 
\[ L \gg (\rho \bb^3)^{-1/2} = (\r \bb)^{-1/2} |\log(\r \aa^2)|\,.
\]
For $L_\a= (\r \bb)^{-1/2} \bb^{-\a}$ as in \eqref{eq:L2d}, this implies $\a> 1$, or equivalently to study Hamiltonians of the form \eqref{eq:HN-kappa2d} for $\kappa>2/3$.   Note that, analogously to the three dimensional case, to recover the correct leading order term in \eqref{eq:LHY2d} from $E_2^{(D)}(N,L)$, one has to choose $L$ larger than the healing length, which is $(\r \bb)^{-1/2}$ in two dimensions.

\medskip

Note that in the regimes described by \eqref{eq:HN-kappa2d} the number of excitations is of the order $N \bb \sim N^{\kappa} \ll N$ for all $\kappa\in [0,1)$. On the other hand, the number of triples needed to recover \eqref{eq:LHY2d} (thinking to a trial state similar to the one in \cite{BCS}) is $N^{-2+2\kappa}$ which is of smaller order with respect to $N$ for all $\kappa \in [0,1)$. This is a key difference with respect to the three dimensional case, where the number of triples increases with $N$ if $\kappa \in (4/9, 2/3)$, and one needs $\kappa>1/2$ to obtain a second order upper bound in the thermodynamic limit.
A similar difference between the $3d$ and $2d$ cases also shows up if one considers trial states of the form used in \cite{BCOPS, FGJMO}, where the Jastrow factor \eqref{eq:jastrow} is multiplied by a quasi free state. Indeed, as discussed in  \cite[Paragraph below (2.7)]{FGJMO}, in two dimensions the contribution coming from the Jastrow factor can be expanded for all $\kappa<1$. As a matter of fact, to match the correct energy, correlations described by the Jastrow factor have to be put on a scale $\ell$ so that $R \ll \ell \ll (\r \bb)^{-1/2}$. Since for any $\kappa \in [0,1)$ we have that $R$ is exponentially small in $N$, while $(\r \bb)^{-1/2} \sim N^\kappa$ there is plenty of space to choose $\ell$ small enough, and hence to make the expansion for the Jastrow factor possible. This is not the case in three dimensions, where correlations described by the Jastrow factor have to be put at a distance $\ell$ satisfying $N^{-1+\kappa} \ll \ell \ll N^{\kappa/2}$.
Indeed, proving a second order upper bound matching \eqref{eq:LHY} in the thermodynamic limit and valid for the hard sphere potential is a current challenge to be overcome.

\bigskip

{\it Acknowledgements.} We acknowledge the support of Istituto Nazionale di Alta Matematica ``F. Severi", through the Intensive Period ``INdAM Quantum Meetings (IQM22)". G.B. acknowledges support through the project ``Progetto Giovani GNFM 2020: Emergent Features in Quantum Bosonic Theories and Semiclassical Analysis". C.C. gratefully acknowledges the support from the European Research Council through the  ERC-AdG CLaQS.

\end{document}